\documentclass[conference]{IEEEtran}

\IEEEoverridecommandlockouts

\usepackage[T1]{fontenc}
\usepackage[utf8]{inputenc}
\usepackage{amsmath,amssymb,amsfonts,amsthm}
\usepackage{bm}
\usepackage{graphicx}
\usepackage{float}
\usepackage{booktabs}
\usepackage{tabularx}
\usepackage{multirow}
\usepackage{xcolor}
\usepackage{cite}
\usepackage{url}
\usepackage{algorithm}
\usepackage{algpseudocode}
\usepackage{array}
\usepackage{enumitem}
\usepackage{balance}

\newcommand{\sys}{\textsc{TRE}}
\newcommand{\metric}{\textsc{TSS}}

\title{Cross-Model Autoscaling for Shared LLM Serving}
\author{
\IEEEauthorblockN{Xin Zhang, Xianyan Xie, Zhen He, Xijin Yin, Xingtong Lin, Bangbo Liang,\\
Zequn Cheng, Peihao Huang, Guo Chen$^{\ast}$\thanks{$^{\ast}$Guo Chen is the corresponding author.}}
\IEEEauthorblockA{Hunan University}
}

\begin{document}
\maketitle

\begin{abstract}
Multi-model LLM serving is moving toward shared MaaS clusters, where co-hosted models compete for a fixed GPU budget while each model experiences time-varying demand and must satisfy its own latency SLO. Existing LLM autoscalers remain largely model-local: their signals expose local runtime activity or delayed latency outcomes, and their scale-up or provisioning decisions do not directly determine how shared capacity should be allocated across competing models.

We present the Token-service-share Rebalancing Engine (TRE), a control-plane framework for hot-switched multi-model LLM serving. TRE introduces Token Service Share (TSS), a calibrated, demand-normalized signal that estimates effective token service per active or queued request and yields a comparable health score across heterogeneous models and SLO classes. Guided by TSS, TRE coordinates bounded receiver--donor capacity movement under a fixed GPU budget: it separates fast rescue from slower rebalancing and incrementally reallocates active replicas toward models with the largest calibrated service deficits.

We implement TRE on a Kubernetes-based hot-switch serving stack without modifying the inference scheduler. Across seven LLM serving traces, TRE reduces P95 end-to-end latency by 11.9--79.0\% and P99 latency by 12.5--72.6\% compared with a state-of-the-art KV-cache-based reactive autoscaler running on the same hot-switch runtime. The gains hold on both targeted stress probes and production-derived conversation/code traces, where TRE reduces P95/P99 latency by 50.8/63.7\% and 79.0/72.6\%, respectively. These results show that effective hot-switched autoscaling requires not only fast replica actuation, but also calibrated service-deficit signals and coordinated cross-model capacity arbitration. Our code and artifacts are available at \url{https://github.com/zxzx9898/Token-service-share_Rebalancing_Engine}.

\end{abstract}


\section{Introduction}
\label{sec:introduction}
Online LLM serving is moving from single-model clusters to shared Model-as-a-Service (MaaS) platforms, where a fixed GPU pool hosts multiple heterogeneous models, each with time-varying demand and its own latency service-level objective (SLO)~\cite{duan2024muxserve,li2023alpaserve,choi2022serving,yu2025prism,patke2024queue}. This shared setting makes autoscaling no longer a per-model sizing problem: when the cluster is fully occupied, scaling up one model requires reclaiming capacity from another. \textbf{The fundamental objective is therefore maximizing overall SLO attainment under a fixed GPU budget---a global optimization that existing autoscalers cannot perform because they are designed to be model-local.}

Reactive systems like Llumnix~\cite{sun2024llumnix} trigger scaling from local runtime pressure (e.g., KV-cache utilization, queue length), while prediction-based systems like PreServe~\cite{jiang2025hierarchical} provision replicas independently for each model's forecasted load. 
These approaches work when idle GPUs are available, but under contention, multiple models may simultaneously demand more resources with no mechanism to arbitrate which deficit is more urgent. \textbf{The root cause is the lack of a cross-model comparable service-deficit signal}: queue length, throughput, and cache-side metrics each expose only a local symptom and do not preserve a stable ordering of service health across heterogeneous models (\S\ref{sec:signal_insufficiency}). Without such a signal, the controller cannot rank which model most needs capacity in service of the global SLO objective.

Realizing global SLO-aware autoscaling raises two core challenges. First, how can the controller construct a unified deficit ranking across models with different sizes, SLO classes, and prefill/decode mixtures? Raw token throughput and queue length are not directly comparable: a 7B model's tokens per second and a 70B model's tokens per second represent different effective service, and an interactive SLO class demands a different service share than a best-effort class. A small model can report high throughput while being deeply under-served, or a large model can show low queue pressure while its prefill backlog silently degrades TTFT. Second, even with a comparable deficit ranking, how should the controller execute cross-model capacity reallocation under a fixed budget? It must choose a receiver, identify an eligible donor, and pace each transfer so that capacity moves respond to persistent service deficits rather than short-lived fluctuations.

We present the Token-service-share Rebalancing Engine (\sys{}), a control-plane framework for hot-switched multi-model LLM serving. To address the first challenge, we introduce Token Service Share (\metric{}), which estimates the effective token service available per outstanding request and normalizes it against a model-specific healthy boundary calibrated from its SLO. \textbf{This normalization is the key: by expressing each model's service state as a fraction of its own minimum healthy requirement, \metric{} places heterogeneous models on a single comparable scale}---a score of 0.5 means the model is receiving half the per-request service it needs to meet its SLO, regardless of whether it is a 7B chat model or a 70B code model.

To address the second challenge, \sys{} coordinates receiver--donor capacity movement through two mechanisms: (1) dual-speed control, which separates fast rescue for critical SLO risk from slower deficit-guided rebalancing; and (2) bounded pairwise transfer, which incrementally moves capacity from a selected donor to the most under-served receiver. Together, these mechanisms determine when reallocation is needed and limit how quickly capacity moves across models.

We implement \sys{} on a Kubernetes-based hot-switch serving stack (AIBrix + vLLM) without modifying the inference scheduler~\cite{tre-codebase}. Across seven LLM serving traces, \sys{} reduces P95 end-to-end latency by 11.9--79.0\% and P99 latency by 12.5--72.6\% versus a state-of-the-art KV-cache-based reactive autoscaler. On production-derived conversation and code traces, the gains reach 50.8/63.7\% and 79.0/72.6\% (P95/P99), respectively.

This paper makes the following contributions:
\begin{itemize}
    \item We propose \metric{}, a demand-normalized service-deficit signal that enables cross-model comparison of effective token service, providing the missing abstraction for global SLO-aware autoscaling.

    \item We design \sys{}, a \metric{}-guided control framework that combines dual-speed control with bounded pairwise reallocation to coordinate cross-model capacity movement under a fixed GPU budget.

    \item We evaluate \sys{} on a hot-switched serving stack, demonstrating consistent tail-latency improvements across synthetic stress probes and production-derived workloads.
\end{itemize}

\section{Background and Motivation}
\label{sec:background}

\subsection{SLOs, Token-Level Demand, and Hot Switching}
\label{sec:llm-serving-dynamics}

\textit{Online LLM serving objectives:} 
In shared LLM serving platforms, SLOs specify latency bounds that inference requests must meet. Common SLO metrics include time-to-first-token (TTFT), time-per-output-token (TPOT), and end-to-end (E2E) latency, typically enforced as tail percentiles (e.g., P95 TTFT $\leq X$, P95 TPOT $\leq Y$)~\cite{zhong2024distserve}. Each hosted model may have its own SLO class: an interactive chat model demands low TTFT, while a code-generation model may tolerate higher TTFT but require low TPOT. When these models share a fixed GPU pool, the cluster-wide objective is to maximize overall SLO attainment by reducing violations of each model's latency bounds rather than optimizing any model in isolation.

\textit{Token-level demand variability:}
In LLM inference, the computational cost of a request is not fixed at arrival; it is determined by the token-level service the request ultimately consumes. A request spans two distinct phases: the prefill phase processes input prompt tokens mostly in parallel, while the decode phase generates output tokens autoregressively. The output length---and thus the total decode work---is unknown until generation completes. Furthermore, the runtime maintains a KV cache for ongoing sequences, and multiplexes requests through continuous batching and chunked prefill~\cite{yu2022orca,kwon2023efficient,agrawal2024taming}. Continuous batching and chunked prefill further couple requests inside a replica. Consequently, two windows with similar arrival rates or request counts can impose very different demand on the same replica set: one may contain short decode sequences, while another may contain long prompts or a growing waiting population. Autoscaling must therefore reason jointly about token-level service progress and the active and waiting requests competing for that service, rather than using request count or arrival rate as a sufficient load proxy.

\textit{Hot switching:}
Cold replica startup requires container startup, runtime initialization, model-weight loading, memory planning, and execution-state warmup, which is too slow for burst-scale control on large models. Hot switching avoids this full cold-start path. It keeps a previously initialized replica in a warm-sleep state and later reactivates it by restoring reusable runtime state and reloading evicted weights. In this paper, hot switching is the mechanism that makes second-scale changes in routable serving capacity practical~\cite{zeng2025medusa,stoyanov2025engine}.

\subsection{Why Existing Signals Fail as Cross-Model Control Signals}
\label{sec:signal_insufficiency}

In a hot-switched multi-model cluster, the controller must decide where active serving capacity should move before tail-latency violations have accumulated. This decision requires a state that ranks the urgency of SLO violation across heterogeneous models.

\textit{Local runtime signals do not preserve cross-model ordering:} Queue length, throughput, KV-cache utilization, and GPU utilization are useful diagnostics, but each exposes only one runtime symptom. None directly states whether a model's active and waiting requests are receiving sufficient effective service, nor whether that model is more urgently under-served than another. A small model can report high throughput while being deeply under-served, or a large model can show low queue pressure while its prefill backlog silently degrades TTFT. Raw token throughput and queue length are not directly comparable across models: a 7B model's tokens per second and a 70B model's tokens per second represent different effective service, and an interactive SLO class demands a different service share than a best-effort class.

\begin{figure}[t]
\centering
\includegraphics[width=0.92\columnwidth]{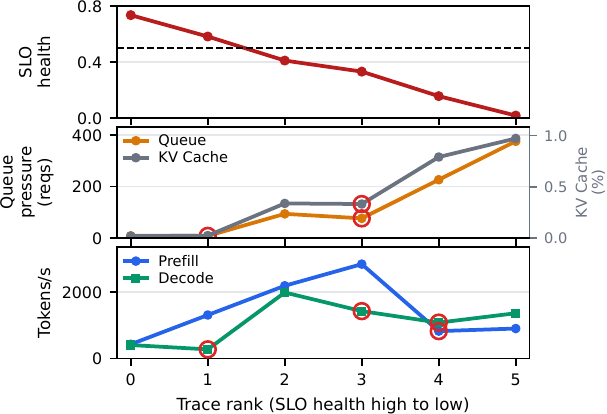}
\vspace{-5pt}
\caption{Fixed-capacity trace segments ordered from healthier to more degraded SLO states. Queue pressure denotes the running-plus-waiting request population, and KV-cache pressure denotes raw cache usage. Queue population, cache pressure, and prefill/decode throughput do not change monotonically with the SLO-derived health ordering; red circles mark representative non-monotone points. These local signals change non-monotonically because they expose different runtime symptoms rather than a common service state.}
\label{fig:ch2_longitudinal_trace_signals}
\vspace{-6pt}
\end{figure}

We test this requirement using fixed-capacity trace segments that vary load intensity and prefill/decode mixture while excluding controller feedback. We order the segments by SLO-derived health, from healthier to more degraded service. A signal suitable for cross-model allocation should preserve this ordering: a model ranked as more urgent should face greater SLO-violation pressure. Figure~\ref{fig:ch2_longitudinal_trace_signals} shows that the running-plus-waiting queue population, KV-cache pressure, and prefill/decode throughput all vary non-monotonically with this ordering. The red open circles mark points where a local signal disagrees with the health ordering, showing that these metrics cannot by themselves produce a reliable health ranking. In one prefill-heavy/decode-heavy pair, the prefill-heavy case violates the target SLO while the decode-heavy case remains healthy, yet their raw running-plus-waiting queues are similar (40.6 vs.\ 42.4 requests). The waiting-only raw inverse is 1.0 in every window for both traces, which only proves that the original average waiting queue is at most one request under the signal's floor, not that it is exactly one. Their KV-cache pressure is also close in the raw trace (0.236\% vs.\ 0.121\%). The degraded prefill-heavy case reports higher prefill TPS (2852.1 vs.\ 831.5 tokens/s) and lower decode TPS (1002.5 vs.\ 1540.0 tokens/s), illustrating why aggregate token progress alone obscures the workload regime. These raw signals cannot serve as a shared scale for comparing which model's SLO violation is more urgent.

\textit{Latency metrics are outcomes, not control states:}
TTFT, TPOT, and tail E2E latency are the ultimate objectives that the system seeks to bound. However, they are observed only after requests have waited, received tokens, or completed service. Latency therefore confirms that degradation has occurred, but it is a poor early state for preemptive resource reallocation.

\begin{figure}[t]
\centering
\includegraphics[width=0.72\columnwidth]{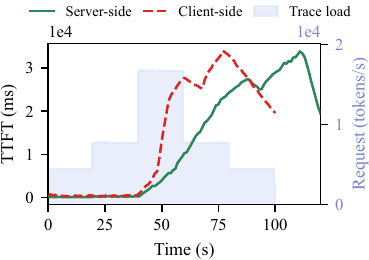}
\vspace{-8pt}
\caption{Latency is a delayed outcome under a controlled prefill-pressure step. The retrospective client-side TTFT attributes delay to request arrival time, while the online server-side metric observes TTFT only after first-token service occurs.}
\label{fig:ch2_latency_lag}
\vspace{-6pt}
\end{figure}

To make this timing issue concrete, Figure~\ref{fig:ch2_latency_lag} compares two TTFT views under a controlled prefill-pressure step. The dashed line is a retrospective client-side view: requests are bucketed by their arrival second, but each request's TTFT is known only after the first token is returned. The solid line is the online server-side metric view: the vLLM TTFT histogram receives a sample only when the first-token event occurs. The server-side view therefore rises after affected requests have already begun waiting and remains elevated while delayed observations drain during recovery. Latency remains essential for SLO accounting and for evaluating control outcomes, but it is too delayed to serve as the primary state for proactive cross-model allocation. The controller instead needs an earlier state that couples current token service with active and waiting demand, which Section~\ref{sec:tss} instantiates as \metric{}.

\subsection{Why Fixed-Budget Reallocation Requires Coordinated Control}
\label{sec:limit_systems}
A comparable cross-model signal is necessary but not sufficient. When no idle GPU is available, adding routable capacity to one model requires capacity to move from another. Autoscaling under a fixed GPU budget is therefore not a collection of independent per-model replica-sizing decisions; it is a shared capacity-allocation problem. The controller must compare service deficits across models, select a receiver and a donor, and limit the amount and pace of each transfer.

\textit{Model-local autoscaling does not resolve contention:}
Reactive systems such as Llumnix trigger scaling from local runtime pressure, while prediction-based systems such as PreServe provision replicas independently from forecasted load~\cite{sun2024llumnix,jiang2025hierarchical}. These approaches can add capacity when idle GPUs exist, but they do not coordinate a shared allocation decision when the cluster is fully occupied. Under contention, multiple models may simultaneously request more capacity. Local signals and per-model forecasts do not determine which deficit should take priority, which lower-pressure model should yield capacity, or how quickly capacity should move between the two.

\textit{The core control challenge:}
Fixed-budget reallocation therefore links two decisions: the controller must rank relative service deficits across heterogeneous models and translate that ranking into bounded receiver--donor capacity movement. The first decision requires a comparable service-deficit signal, which \metric{} provides in Section~\ref{sec:tss}. The second requires a coordinated control framework, which \sys{} provides in Section~\ref{sec:tre}, to separate urgent rescue from slower rebalancing and to move capacity incrementally under a fixed GPU budget.

Together, Sections~\ref{sec:signal_insufficiency} and~\ref{sec:limit_systems} define the two linked requirements for global SLO-aware autoscaling: a comparable service-deficit signal across heterogeneous models, provided by \metric{} in Section~\ref{sec:tss}, and a coordinated control framework, provided by \sys{} in Section~\ref{sec:tre}, that translates this ranking into bounded receiver--donor capacity reallocation under a fixed GPU budget.

\section{Token Service Share: A Demand-Normalized Service State}
\label{sec:tss}
Section~\ref{sec:background} showed that existing online signals fail in four ways: queue and throughput decouple supply from demand; cache and GPU metrics expose only local resource symptoms; latency is the correct objective but arrives too late as a feedback signal; and none of these signals is cross-model comparable because the same token rate can represent different effective service for different models. A usable control signal must therefore (i) couple token-level supply with per-request demand, (ii) be computable before violations accumulate, and (iii) normalize onto a common scale so that a controller can rank heterogeneous models by a single urgency score.

We introduce \emph{Token Service Share} (\metric{}) for this purpose. \metric{} is not a per-request latency predictor, nor a direct SLO measurement. It is a control-side proxy that estimates the effective token service available to each request that is either being served or already waiting at the front door. Its role is to provide a leading, comparable, and actionable service-deficit ranking.

\subsection{Design Rationale and Definition}
\label{sec:tss-definition}
\textit{Why a ratio, not a raw metric:}
Aggregate throughput can remain high even when it is divided among too many requests, while queue length does not reveal how quickly that backlog is being served. What matters for SLO health is the ratio of effective token service to the population of requests competing for it. This directly addresses the supply--demand decoupling identified in Section~\ref{sec:signal_insufficiency}.

For model $m$, let $A_m(t)$ be the number of active requests receiving runtime service, and let $W_m(t)$ be the number of waiting requests at the model's front-door queue. Over a monitoring window of width $\Delta$, the controller observes prefill and decode token counts. We write $r^{\mathrm{prefill}}_m(t,\Delta)$ and $r^{\mathrm{decode}}_m(t,\Delta)$ for the corresponding token-service rates, computed as the observed token counts divided by $\Delta$. The complete raw \metric{} ratio is
\begin{equation}
\mathrm{TSS}^{\mathrm{raw}}_m(t)
=
\frac{
w_p r^{\mathrm{prefill}}_m(t,\Delta)
+
r^{\mathrm{decode}}_m(t,\Delta)
}{
A_m(t) + w_q W_m(t)
}.
\label{eq:tss_raw}
\end{equation}

\textit{The numerator: effective token service:}
Prefill and decode are not interchangeable: prefill is parallel computation that determines TTFT, whereas decode is sequential generation that determines TPOT. Treating them as a single effective-service unit is a deliberate engineering simplification: \metric{} targets cross-model ranking of service deficit rather than precise per-request latency prediction, and a linear aggregation with the model-specific weight $w_p$ is sufficient for this control purpose while remaining online-computable. Because prefill can process many prompt tokens in parallel, one prefill token generally contributes less to per-request SLO health than one decode token. Accordingly, the calibrated prefill weight $w_p$ is generally less than one. Its value is constrained by the model's prefill parallelism and the SLO class's relative sensitivity to TTFT versus TPOT, rather than fitted as an open-ended degree of freedom.

\textit{The denominator: effective outstanding demand:}
Active requests are counted at full weight because they already occupy execution slots and memory bandwidth. Waiting requests do not yet consume runtime resources, but they represent committed demand: their latency budgets are being spent before they receive token service, and they must still be admitted and executed. The calibrated waiting weight \(w_q\), generally above one, maps this admission-side risk into the same outstanding-demand unit. It does not imply that waiting requests directly slow the current iteration; it makes queued work visible before its latency impact appears in telemetry.

Eq.~\eqref{eq:tss_raw} is therefore the service-share state used by the controller. A lower value means that less effective token service is available per active or waiting request, indicating a larger model-level service deficit. A model can report higher aggregate throughput but lower \metric{} when that throughput is shared by a sufficiently larger active and waiting population.

\textit{Why this ratio preserves the SLO-health ordering that aggregate throughput lacks:}
The prefill-heavy/decode-heavy pair in Figure~\ref{fig:ch2_longitudinal_trace_signals} illustrates why this normalization is needed. Aggregate TPS ranks the pair in the wrong direction because parallel prefill dominates raw token progress. \metric{} avoids this inversion by discounting prefill progress according to its lower per-request SLO contribution and normalizing the resulting effective service by the urgency-weighted outstanding population. The prefill-heavy case consequently receives a lower \metric{} than the healthy decode-heavy case, consistent with its observed SLO violation. Within a fixed model and SLO class, decreasing \metric{} indicates increasing service deficit and a higher risk of tail-latency violations.

\textit{Scope and limitation:}
\metric{} is a proxy that assumes ordinary request-level heterogeneity is absorbed by the runtime's continuous batching policy. Extreme length skew or physically disaggregated prefill/decode resources may require a stage-aware service state; Section~\ref{sec:discussion} discusses this boundary.

\subsection{Smoothing, Normalization, and Calibration}
\label{sec:tss-calibration}
Continuous batching, chunked prefill, and decode completion can create short-window token-service spikes. The controller therefore applies exponential smoothing with parameter $\alpha$: each window's \metric{} combines the current raw ratio with the previous smoothed value, avoiding transient oscillations while preserving control-loop reactivity.

Raw \metric{} is model-local. Calibration defines a healthy boundary $\theta_m$ for each model and SLO class:
\begin{equation}
Z_m(t)
=
\frac{\mathrm{TSS}_m(t)}{\theta_m}.
\label{eq:tss_normalized}
\end{equation}
$Z_m(t)$ measures service share relative to the model's own healthy requirement: values above one indicate service above the calibrated requirement, whereas values below one indicate service below it. This directly addresses the cross-model comparability problem identified in Section~\ref{sec:signal_insufficiency}.
\textit{Calibration:}
The profile is derived from offline profiling across the healthy-to-degraded transition. The parameters are constrained by physical and control-side invariants: $w_p$ by prefill parallelism and observed TTFT--TPOT tradeoffs; $w_q$ by the urgency of waiting work under the target SLO class; $\alpha$ by the control-loop reaction time; and $\theta_m$ by the smallest smoothed \metric{} that achieves target SLO reliability. The profile contains only six scalars per model--SLO pair and is loaded before online execution. Recalibration is a slow-path maintenance action, not a per-tick fitting procedure.

The controller partitions $Z_m(t)$ with hysteresis thresholds $\tau_{\mathrm{crit}} < 1 < \tau_{\mathrm{surplus}}$: critical identifies receiver-side pressure, surplus identifies preferred donor candidates, and nominal provides a stable region around the healthy boundary. These regions stabilize receiver--donor selection by preventing small fluctuations around $Z_m(t)=1$ from triggering a capacity transition.

The control profile is derived offline for each model--SLO pair across the healthy-to-degraded transition. The parameters are constrained by physical and control-side semantics: \(w_p\) reflects prefill parallelism and the observed TTFT--TPOT tradeoff; \(w_q\) reflects the urgency of waiting work under the target SLO class; \(\alpha\) reflects the desired control-loop reaction time; and \(\theta_m\) is the smallest smoothed \metric{} value consistent with target SLO reliability. The profile contains six scalars per model--SLO pair and is loaded before online execution; it is not re-fitted at each control tick.

\section{\sys{}: \metric{}-Guided Cross-Model Autoscaling}
\label{sec:tre}

\begin{figure}[t]
\centering
\includegraphics[width=\columnwidth]{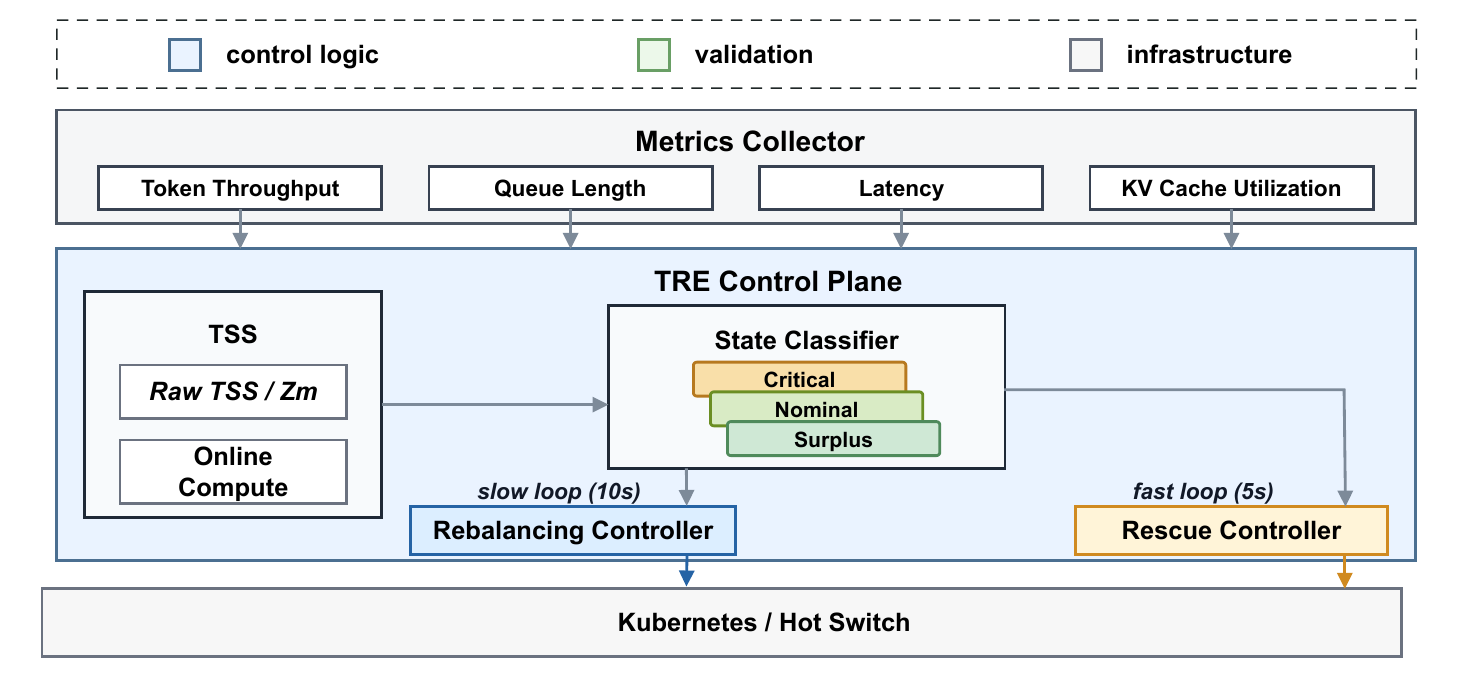}
\vspace{-8pt}
\caption{Overview of the \sys{} control plane. The signal layer derives normalized service states \(Z_m(t)\) and classifies models as critical, nominal, or surplus. The control layer uses a fast rescue path and a slower rebalancing path to issue bounded routable-capacity transitions. The actuation layer applies local donor-side guards before committing a capacity release.}
\label{fig:tre-architecture}
\vspace{-10pt}
\end{figure}

\sys{} is a control-plane framework that translates cross-model service-deficit rankings into bounded capacity reallocation. It has two core control mechanisms: dual-speed control separates emergency rescue from routine rebalancing, and bounded pairwise transfer incrementally moves capacity without oscillation. Before a selected donor-side release is committed, the actuation layer applies local guardrails; these checks can defer or reject the release but do not determine global allocation.

\subsection{Overview and Control Objective}
\label{sec:tre-interface}

\sys{} aims to maximize overall SLO attainment under a fixed GPU budget. When idle GPUs exist, this reduces to independent per-model scaling, as in local reactive or forecast-based provisioning approaches~\cite{sun2024llumnix,jiang2025hierarchical}. Under contention, the problem becomes cross-model arbitration: increasing one model's routable capacity requires another model to release it. This is approximately a calibrated max--min service-share problem: capacity should move toward the model with the lowest normalized service share, subject to fixed-budget feasibility, minimum routable-replica floors, and local donor-side guardrails.

The two core mechanisms below address this objective at different time scales. Dual-speed control (Section~\ref{sec:tre-dual-speed}) determines when intervention is needed, while bounded pairwise transfer (Section~\ref{sec:tre-bounded-transfer}) selects a receiver--donor transition and limits its pace. \metric{} provides the shared signal for these allocation decisions. KV-cache utilization and tail latency remain local guardrails that can exclude a candidate donor release; they do not determine which model receives capacity or the cross-model direction of reallocation.

\subsection{Dual-Speed Control}
\label{sec:tre-dual-speed}
A single control cadence cannot serve both emergency rescue and routine rebalancing. A cadence tuned for rescue would let routine rebalancing chase short-lived fluctuations and increase replica churn, whereas a cadence tuned for stability would leave critical deficits unresolved for multiple service intervals. \sys{} therefore separates a fast rescue loop from a slower deficit-guided rebalancing loop.

\textit{Rescue:}
When a model enters the critical region, breaches a latency guardrail, or loses routable capacity, \sys{} enters the fast rescue path. Rescue selects the most urgent receiver: the model with the lowest $Z_m(t)$ among critical models, while hard-guard violations preempt all other events. It acquires capacity in least-disruptive order: first idle hot-switch capacity, then non-routable warm replicas, and only then donor-side release. This ordering minimizes the probability of triggering unsafe donor shrink during an emergency. The fast loop runs at a 5\,s cadence in our prototype.

\textit{Deficit-guided rebalancing:}
When no rescue is pending, the controller runs the slower rebalancing loop at a 10\,s cadence. It selects the receiver as
\begin{equation}
m_{\mathrm{recv}} =
\arg\min_m Z_m(t),
\label{eq:tre-receiver}
\end{equation}
i.e., the model with the largest calibrated service deficit. It then searches for an eligible donor, preferring surplus models and considering nominal models only when no surplus or idle capacity exists. A candidate donor is released only if its local guardrails permit a routable-capacity transition at actuation time.

The two loops are mutually exclusive: when rescue is pending, the slow loop defers all donor-side operations to avoid reclaiming capacity from a model that is itself under pressure. Conversely, a rebalancing transfer in progress is not preempted by a new rescue request unless the target model breaches a hard guardrail.

\subsection{Bounded Pairwise Transfer}
\label{sec:tre-bounded-transfer}
Even after selecting a receiver and a donor, \sys{} does not re-solve a global allocation problem at every control tick. Workload conditions can change before a multi-model plan is fully enacted, and hot-switch actuation is inherently discrete. \sys{} therefore uses a greedy, quantized pairwise rule.

The normalized deficit determines the direction of movement rather than a fractional replica count. For a selected receiver--donor pair, the controller issues at most one GPU-budget-feasible routable-capacity transition per tick. A transition is considered only when the receiver is below its healthy boundary, the donor remains above its minimum routable-replica floor, local donor-side guardrails permit the release, and the GPU footprints of the released and activated capacity satisfy the fixed cluster budget. The transition is therefore not assumed to be a one-to-one replica exchange across heterogeneous models.

After each transition, the controller observes the resulting $Z_m(t)$ values and re-evaluates at the next tick. This bounded rule prevents oscillatory corrections while preserving the max--min direction. This small-step policy limits corrective churn and reduces the risk of overshoot, without claiming a closed-form global optimum or formal convergence under changing demand.

\textit{Donor-side release guard:}
\label{sec:donor-release-guard}
The donor-side release guard does not rank models or choose a receiver. For a selected donor-side release, the adapter enforces the minimum-routable-replica guard, hides the candidate from new-request routing while in-flight requests drain, and observes local latency, KV-cache, and donor-health conditions. If a guardrail is violated, the adapter restores routing and defers the transition; otherwise, it commits the release by sleeping the replica. This guard constrains the actuation of a selected transfer rather than defining a separate control policy.

\section{Implementation}
\label{sec:implementation}

We implement \sys{} as a Kubernetes-based control-plane service on top of an AIBrix serving stack and a vLLM hot-switch runtime. AIBrix provides the serving and gateway layer, while vLLM provides the inference runtime and sleep/wake primitives. \sys{} does not modify the intra-replica inference scheduler: it does not change vLLM's continuous batching, prefill/decode scheduling, or token execution path. Instead, it operates at the serving-layer routing and replica-lifecycle level.

\subsection{System Boundary and Replica State}
\label{sec:impl_replica_model}
The prototype has two components. The \sys{} controller consumes per-model telemetry, applies the calibrated \metric{} profile, classifies each model into a service region, and issues bounded requests for routable-capacity transitions. A service-manager adapter translates these requests into AIBrix routing changes and vLLM sleep/wake calls. This split keeps the serving stack responsible for request execution, while \sys{} controls model-level routable capacity and the adapter enforces local donor-side release guards during actuation.

The control unit is a serving replica, implemented as a Kubernetes Pod running a vLLM process for one model. A replica may use one or more GPUs depending on the model-parallel placement configured by the serving stack; \sys{} does not change this internal placement. The controller distinguishes created replicas from routable replicas. A created replica may still be hidden, sleeping, or reactivating, and therefore is not counted as immediate serving capacity. Only runtime-ready and gateway-visible replicas are counted as routable.

\sys{} follows the lifecycle concepts used in the design. An active replica is routable and can receive new requests. A hidden replica is still awake but excluded from new-request routing; in the adapter this is implemented with a custom gateway-consumed Pod annotation, \texttt{aibrix.ai/route-hidden}. A sleeping replica remains hidden and has been committed to vLLM sleep, so it acts as warm standby. A reactivating replica is being woken and becomes routable only after runtime readiness and gateway visibility are restored.

Sleeping releases GPU memory in the vLLM runtime but does not release the Kubernetes Pod or its scheduler-level GPU allocation. The service-manager adapter therefore performs hot-switch capacity accounting above Kubernetes. It treats GPUs whose assigned replicas are all sleeping as available for hot-switch capacity planning, while \sys{} continues to reason about routable capacity rather than raw Kubernetes GPU requests alone.

\subsection{Telemetry and Controller Execution}
\label{sec:impl_telemetry}

The prototype exports per-Pod vLLM metrics to a Redis-backed store, and the controller aggregates them into per-model control-window observations. Prompt-token and decode-token counters are converted into window deltas for \metric{} computation. Running and waiting request counts represent outstanding demand. KV-cache utilization, swapped-request counts, and latency histograms are retained as guard and diagnostic telemetry. Replica counts and routing visibility are obtained from the service-manager adapter because runtime metrics alone do not expose lifecycle state.

Each model has a static control profile generated by the slow-path calibration procedure in Section~\ref{sec:tss-calibration}. The profile contains the quantities used by the online \metric{} computation: \(w_p\), \(w_q\), \(\alpha\), \(\theta_m\), and the critical/surplus thresholds. The controller does not re-fit these parameters in the fast loop; it only applies the loaded profile to fresh telemetry, updates \(Z_m\), and classifies the model into the critical, nominal, or surplus region.

The controller uses one main polling loop with two logical cadences. Each tick refreshes replica state, collects metrics, updates \metric{}, and emits bounded transition requests. The fast cadence handles rescue for critical models or hard-guard violations. The slower cadence performs routine rebalancing when no rescue action is pending. Requests for additional routable replicas are sent to the adapter. Requests that reduce routable capacity are subject to local donor-side release guards, which may defer or reject a selected transition. KV-cache utilization and tail latency are not merged into \metric{}; they remain local guardrails that can veto a donor release, but do not change the cross-model allocation ranking.

\subsection{Gateway Integration and Donor-Side Release Guard}
\label{sec:impl_gateway_hotswitch}
The service-manager adapter exposes a small set of lifecycle operations. To add routable capacity, it wakes a hidden sleeping replica when one is available, checks that reactivation does not conflict with GPU-memory availability, invokes vLLM wake-up, and restores gateway visibility only after the runtime is ready.

For routable-capacity release, the adapter enforces the local donor-side guard introduced in Section~\ref{sec:tre-bounded-transfer}. It first enforces the minimum-routable-replica floor, then hides the selected replica from new-request routing through the gateway-consumed \texttt{aibrix.ai/route-hidden} annotation. This routing-only transition keeps the vLLM process awake while in-flight requests drain during a bounded interval. The adapter observes local donor-side conditions, including latency, queue growth, and KV-cache state. If a guardrail is violated, it restores gateway visibility and defers the release; otherwise, it commits the transition by invoking vLLM sleep. These checks constrain the actuation of a selected donor release but do not recompute the global allocation decision.

If a committed sleep interrupts unfinished work after the drain interval, the current prototype recovers it through reissue rather than cross-replica KV-cache migration. This path may recompute part of the prefill work and is a prototype limitation discussed in Section~\ref{sec:discussion}.

\section{Evaluation}
\label{sec:evaluation}

We evaluate \sys{} through three research questions that address the validity of its cross-model control signal, its end-to-end effectiveness, and its deployment cost.

\begin{itemize}[noitemsep,topsep=0pt,leftmargin=*]
\item \textbf{RQ1:} Does \metric{} provide a correct and cross-model comparable ranking?
\item \textbf{RQ2:} Does \sys{} improve tail latency in end-to-end serving?
\item \textbf{RQ3:} What deployment overhead is introduced by \sys{}'s control loop and hot-switch actuation?
\end{itemize}


\subsection{Experimental Setup}
\label{sec:eval-setup}

\textit{Testbed:} All experiments run on two machines, each equipped with four NVIDIA A100 40\,GB GPUs connected by NVLink. The serving stack is a Kubernetes-based multi-model runtime built on vLLM~\cite{kwon2023efficient} with hot-switch support following the AIBrix design~\cite{team2025aibrix}. We serve three models simultaneously: \texttt{dsllama-8b}, \texttt{dsqwen-7b}, and \texttt{dsqwen-14b}. These models are chosen deliberately: they differ in GPU footprint (one versus two replica-group GPUs), which exercises TRE's cross-model allocation under heterogeneous capacity costs.

\textit{Workloads:} We use seven traces. The first five are \emph{stress probes}, each designed to isolate a specific autoscaling failure mode. \emph{Sinusoidal} tests whether the controller can track gradual, periodic pressure without overreacting. \emph{Decode} sustains a decode-heavy burst to expose TPOT amplification under KV-cache pressure. \emph{Prefill} introduces prompt/decode mix shifts to create per-model stage skew. \emph{Alternating} holds aggregate load constant while rotating the hot model across the cluster, isolating the cross-model transfer decision from any change in total offered load. \emph{Simul-Spike} injects repeated synchronized spikes to stress the emergency rescue path under shared contention. The final two traces, \emph{Real-Conv} and \emph{Real-Code}, are derived from production traffic; they test whether conclusions drawn from the stress probes hold under irregular arrivals and heavy token distributions that were not explicitly targeted by the probe design. Each trace runs for approximately 720\,s.

\textit{Baseline:} Our primary comparison is \textbf{KV-Auto}, a state-of-the-art reactive autoscaler that follows the Llumnix-style design~\cite{sun2024llumnix}: each model is scaled independently when its local KV-cache utilization crosses a fixed threshold, with no cross-model allocation objective. KV-Auto runs on the \emph{identical} serving runtime, request scheduler, and hot-switch infrastructure as TRE---only the control policy differs---so any latency difference is attributable to the autoscaling decision logic rather than to infrastructure variation. We do not include prediction-based systems such as PreServe~\cite{jiang2025hierarchical} as end-to-end comparisons, because their design targets a different operating regime. PreServe's workload predictor uses fixed 10-minute windows and a proactive scaler that rolls predictions forward to the next provisioning window, targeting coarse-grained cold-start planning. We reproduced the released mLSTM predictor on our Azure-code and Azure-conv traces at 5\,s windows---the control interval relevant to hot-switch rebalancing. At 5\,s granularity, mLSTM is no more accurate than simply using the most recent observed load as the next-window prediction (Azure-code MAPE: 17.8\% vs. 18.1\%; Azure-conv: both 10.4\%), so it provides little additional value for seconds-scale rebalancing. 

\textit{Metrics:} Our primary metrics are P95 E2E latency, since autoscaling failures appear as short tail excursions that the mean dilutes; we also report mean E2E latency to confirm tail gains do not hurt average behavior. We decompose results into P95 TTFT and P95 TPOT to separate prefill-side admission pressure from decode-side service quality. Percentiles are computed over successful requests only, with success rate reported separately.

\begin{figure}[!t]
    \centering
    \includegraphics[width=\columnwidth]{%
        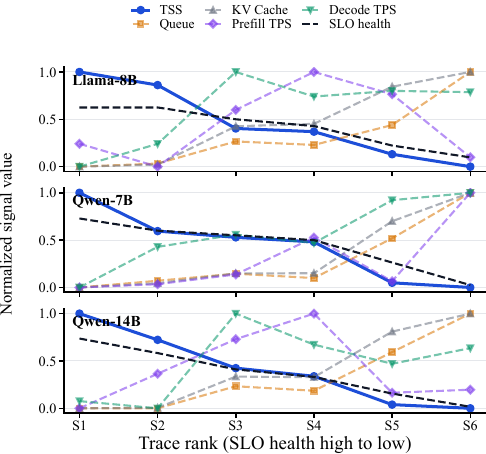}
    \vspace{-12pt}
    \caption{Validating \metric{} against local runtime signals across
    per-model trace ranks. Each panel corresponds to one model. The solid line
    shows min--max normalized \metric{}, translucent dashed lines show local
    runtime signals, and the black dashed line shows SLO health ordered from
    high to low.}
    \label{fig:eval-rq1-tss-validation}
    \vspace{-8pt}
\end{figure}

\begin{figure*}[!t]
    \centering
    \includegraphics[width=\textwidth]{%
        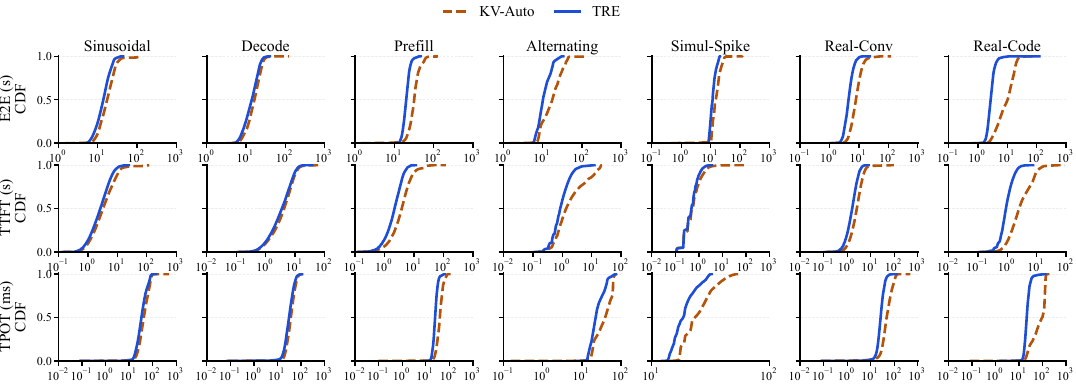}
    \vspace{-6pt}
    \caption{Request-level latency CDFs for \sys{} and KV-Auto
        across all seven traces. Rows show E2E latency, TTFT, and TPOT;
        columns show traces. CDFs are computed over successful requests
        only; all runs have 100.0\%/100.0\% success rate. E2E and TTFT
        use seconds, while TPOT uses milliseconds.}
    \label{fig:eval-latency-cdfs}
    \vspace{-6pt}
\end{figure*}

\begin{figure*}[!t]
    \centering
    \includegraphics[width=0.96\textwidth]{%
        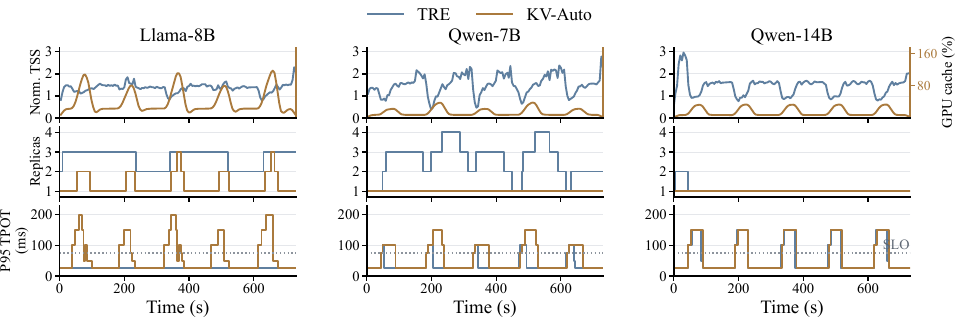}
    \vspace{-6pt}
    \caption{Per-model control dynamics under the Simul-Spike trace.
        Each column corresponds to one served model. Rows show the control
        signal (normalized TSS for \sys{} and GPU-cache pressure for
        KV-Auto), active replicas, and P95 TPOT.}
    \label{fig:eval-simul-spike-models}
    \vspace{-6pt}
\end{figure*}


\subsection{RQ1: TSS Ranking Correctness and Cross-Model Comparability}
\label{sec:eval-rq1}

The first question is whether \metric{} is only a within-model health indicator, or whether it can also support cross-model arbitration. This distinction matters for \sys{} because a capacity-transfer decision compares heterogeneous models: the controller must decide whether, for example, a stressed \texttt{dsqwen-7b} should receive capacity from a less stressed \texttt{dsllama-8b}, even though their raw serving rates and queue dynamics live on different scales.

We evaluate this property using per-model fixed-capacity trace ranks that cover each model's transition from healthy to degraded service under its target SLO profile. The trace ranks are selected within each model's operating range, so the comparison tests whether the same normalized \metric{} construction preserves health ordering despite model-specific workload regimes.

Figure~\ref{fig:eval-rq1-tss-validation} compares normalized \metric{} with local runtime signals over six trace ranks ordered from higher to lower SLO health for each model. Because queue pressure, KV-cache pressure, prefill TPS, decode TPS, and normalized \metric{} have different units before plotting, each curve is min--max scaled within the corresponding model; the dashed curve shows SLO health on the same 0--1 scale. Normalized \metric{} follows the SLO-health ordering most consistently across the three models. In contrast, the other metrics do not maintain a consistent relationship with health ordering: they change direction and flatten different prompt/decode mixtures, so they cannot reliably support health ranking.


\subsection{RQ2: End-to-End Tail Latency}
\label{sec:eval-rq2}

\sys{} converts the cross-model health ranking from TSS into concrete capacity-transfer decisions. We first ask whether those decisions improve end-to-end serving quality rather than merely changing controller internals. Figure~\ref{fig:eval-latency-cdfs} reports the request-level latency distributions across both targeted stress scenarios and production-derived workloads. Across all seven traces, \sys{} reduces P95 E2E latency by 11.9--79.0\% and P99 E2E latency by 12.5--72.6\% relative to KV-Auto. The gains are consistent across targeted stress probes and production-derived workloads: \sys{} outperforms KV-Auto on every trace, with particularly large reductions on Real-Conv (50.8\%/63.7\% P95/P99) and Real-Code (79.0\%/72.6\%). Mean E2E latency also improves by 14.8--73.0\%, indicating that the tail gains do not come from sacrificing average-case behavior.

We next examine Simul-Spike in detail because it simultaneously stresses burst recovery and cross-model capacity arbitration. On this trace, \sys{} reduces aggregate P95/P99 E2E latency from 26.61/30.57\,s to 18.41/19.49\,s. The improvement is not uniform across models: P95 E2E falls by 49.7\% for \texttt{dsllama-8b} and 30.7\% for \texttt{dsqwen-7b}, while \texttt{dsqwen-14b} changes only slightly (2.0\%). Thus, Simul-Spike is mainly a test of whether the controller can identify which synchronized bursts are most recoverable, rather than simply adding capacity to every pressured model.

Figure~\ref{fig:eval-simul-spike-models} shows the corresponding per-model dynamics. KV-Auto reacts only to local cache pressure, so \texttt{dsqwen-7b} and \texttt{dsqwen-14b} remain at one replica and \texttt{dsllama-8b} is scaled reactively after pressure has already accumulated. In contrast, \sys{} uses normalized TSS to move capacity across models: it raises \texttt{dsllama-8b} to three replicas and \texttt{dsqwen-7b} up to four replicas during spike periods. This removes window-level P95 TPOT excursions above the 75\,ms SLO for \texttt{dsllama-8b} (41 windows under KV-Auto versus none under \sys{}) and reduces them for \texttt{dsqwen-7b} (46 to 13 windows). \texttt{dsqwen-14b} remains the hardest case---its spike-window TPOT excursions are comparable under both policies---but its request-level E2E tail does not worsen while the other two models improve substantially.

In summary, \sys{} consistently reduces tail latency across all evaluated traces. The Simul-Spike breakdown shows that the end-to-end gain comes from selective cross-model recovery: TRE prioritizes the models where additional replicas most reduce tail latency, rather than uniformly suppressing every per-model spike.


\subsection{RQ3: Control and Actuation Overhead}
\label{sec:eval-overhead}

We quantify two sources of overhead: the latency cost of hot-switch actuation triggered by \sys{}, and the HBM footprint of sleeping replicas retained as warm standby capacity.

\textit{Actuation latency:} We measure sleep and wake latencies for each model. Normal sleep completes in 1.87\,s on average and wake-up in 1.31\,s on average, both within \sys{}'s 10\,s rebalancing interval. Thus, a capacity transition initiated by the slow rebalancing loop can complete within a single control period in our testbed. The first sleep operation is slower (12.19\,s on average) due to one-time runtime cleanup; subsequent transitions follow the fast warm path.

\textit{Memory footprint of sleeping replicas:} Sleep releases 94.9--97.0\% of per-GPU model memory, leaving only 1.0--1.8\,GB of reactivation metadata resident per sleeping replica. This allows the prototype to retain warm standby capacity with a limited HBM footprint. These numbers characterize actuation cost; controller decision latency is negligible because the \metric{} engine performs only arithmetic over per-window token counts and the rebalancing loop performs a greedy pairwise comparison over at most a few tens of model entries per tick.

Overall, hot-switch actuation fits within the prototype's control intervals, and sleeping replicas consume under 2\,GB of HBM per replica. These costs support the feasibility of fine-grained capacity movement in the evaluated deployment.

\section{Related Work}
\label{sec:related_work}
\textit{Cloud-native service autoscaling:}
Cloud-native autoscalers and SLO-aware managers adjust replica counts or resource allocations using utilization, request demand, latency, or application-tier feedback~\cite{kubernetes-hpa,qiu2020firm,zhang2021sinan,bhardwaj2023cilantro}. These inputs generally estimate pressure from request load, resource use, or observed application behavior. In LLM serving, however, effective demand continues to evolve after admission with prompt and decode work, continuous batching, and KV-cache residency. Rather than treating models as independent scaling targets, \sys{} uses \metric{} to compare current service deficits across co-hosted models and coordinate capacity reallocation when the GPU budget is contended.

\textit{LLM serving runtimes and schedulers:}
LLM serving systems improve multi-model multiplexing and serving efficiency through runtime scheduling, KV-cache management, request admission, fairness, migration, routing, and placement~\cite{li2023alpaserve,duan2024muxserve,team2025aibrix,yu2022orca,kwon2023efficient,agrawal2024taming,patke2024queue,sheng2024fairness,sun2024llumnix,jain2025performance}. These systems optimize how requests and model state are executed within the serving substrate. \sys{} is complementary: it operates at the control plane above the request scheduler and uses a cross-model deficit ranking to determine when routable capacity should move between models under a fixed GPU budget.

\textit{LLM autoscaling signals and provisioning:}
Recent LLM autoscaling work explores reactive signals, such as queueing, utilization, throughput, and token progress~\cite{patke2025hierarchical,li2025taming,lai2025tokenscale}, as well as predictive provisioning based on future workload or resource demand~\cite{jiang2025hierarchical,jaiswal2025sageserve}. These approaches capture local pressure, stage-specific bottlenecks, or future provisioning needs, but do not provide a common cross-model ordering of current service deficit under a fixed GPU budget. \metric{} instead estimates demand-normalized token service at each model's front door, enabling \sys{} to compare under-service and guide replica transfers.

\textit{Disaggregation, hot switching, and GPU sharing:}
Disaggregated and KV-cache-centric serving systems address stage separation, KV-cache transfer, and placement constraints~\cite{zhong2024distserve,patel2024splitwise,qin2024mooncake}. Fast scaling and GPU-sharing systems reduce the cost of launching, migrating, or sharing serving capacity~\cite{zeng2025medusa,stoyanov2025engine,fu2024serverlessllm,zhang2025blitzscale,xiang2025aegaeon,yu2025prism}. These mechanisms are complementary to \sys{}, which uses hot switching for actuation and reallocates routable replicas across co-hosted models under a fixed GPU budget; stage-aware control for disaggregated deployments remains outside its current scope.

\section{Discussion and Limitations}
\label{sec:discussion}
\subsection{Deployment Insights}
\sys{}'s benefit depends on workload conditions and the degree of GPU contention. In our evaluation, larger gains appear when models experience irregular bursts, rapid popularity shifts, and sustained competition for a fixed GPU budget. In these settings, relative service deficits can persist long enough that coordinated reallocation improves on independent model-local reactions.

The benefit is smaller when demand changes gradually and predictably, or when idle GPUs remain available. In those settings, adding capacity does not require a receiver--donor tradeoff, and independent per-model provisioning can be sufficient. Thus, our results position \sys{} as a control mechanism for contended multi-model deployments, rather than a replacement for model-local autoscaling in every serving setting.

\subsection{Limitations and Boundary Conditions}
\textit{Disaggregated serving:}
\sys{}'s replica-level abstraction assumes co-located prefill and decode. Extending it to disaggregated and KV-cache-centric serving systems, such as DistServe, Splitwise, and Mooncake, requires two changes~\cite{zhong2024distserve,patel2024splitwise,qin2024mooncake}. First, \metric{} would need to become stage-aware: the scalar score would become a stage-specific vector that estimates prefill and decode service deficits separately, because physically distinct GPU pools may have independent queueing dynamics. Second, donor-side release guards would need to account for inter-stage coupling: reducing decode capacity must not create KV-cache-transfer or admission pressure that violates prefill-side service. The core concepts---demand-normalized deficit ranking and guarded capacity movement---remain applicable, but the unit of reallocation shifts from a routable replica to a stage-specific resource slice.

\textit{Calibration and operational drift:}
\metric{} is a calibrated proxy, not a queueing-theoretic identity. Its cross-model comparability depends on profiles that remain fixed during online execution. In production MaaS platforms, models are updated frequently and workload distributions shift with user behavior. The current slow-path recalibration assumes that these changes are infrequent relative to the control-loop timescale. A practical deployment would benefit from online adaptation, for example, a Bayesian update of \metric{} using streaming latency feedback, or lightweight canary replay on a fraction of replicas. We leave the exploration of online calibration with bounded perturbation to the live serving path to future work.

\textit{In-flight work during release:}
The donor-side release guard hides a candidate replica and lets in-flight requests drain before invoking sleep. In the common case, no request-level recovery is needed because the replica remains awake during the drain interval. If the bounded interval expires while long generations are still active, the prototype falls back to reissue rather than cross-replica KV-cache migration. This fallback may recompute part of the prefix and can be expensive for long-context workloads. We therefore treat reissue as a prototype boundary, not as part of \sys{}'s allocation policy. Production runtimes with KV-cache transfer could replace it with state migration, or choose between reissue and transfer using remaining generation length, KV-cache size, and interconnect bandwidth. This boundary is orthogonal to \metric{}-guided receiver--donor selection.

\subsection{Future Work}

This work introduces two design elements that are not limited to the current prototype: Token Service Share, a demand-normalized and cross-model comparable service-deficit signal, and a bounded control pattern that translates this ranking into receiver--donor capacity reallocation under a fixed GPU budget. We outline three directions for extending this design beyond the present work.

\textit{Stage-aware service shares for disaggregated serving:}
The most direct extension is to disaggregated prefill/decode serving. Rather than maintaining one model-level state, the controller would maintain separate \metric{} states for prefill and decode, and coordinate stage-specific capacity across their corresponding GPU pools. This extension would expose bottlenecks that a co-located scalar can hide, but it would also require the controller to account for cross-stage queueing, KV-cache-transfer pressure, and placement constraints when selecting a capacity movement.


\textit{Finer-grained capacity movement and recovery:}
Future runtimes may move stage-specific resource slices or routing weights rather than whole routable replicas, and may recover in-flight work through reissue, delayed draining, or KV-cache transfer. Extending \sys{} to these operations would require a capacity model that represents their resource footprint, actuation delay, and recovery cost, while preserving the bounded receiver--donor nature of the control loop. Such a model could choose among recovery paths based on remaining generation length, cache-transfer cost, and user-visible interruption.

\section{Conclusion}
\label{sec:conclusion}

We presented \sys{}, a control plane for cross-model capacity arbitration in shared LLM serving under a fixed GPU budget. \sys{} uses \metric{}, a calibrated token- and queue-aware measure of effective token service per outstanding request, to compare service deficits across heterogeneous models and SLO classes and guide bounded receiver--donor capacity transfers. Implemented on a Kubernetes-based AIBrix and vLLM stack without modifying the inference scheduler, \sys{} reduces P95 end-to-end latency by 11.9--79.0\% and P99 latency by 12.5--72.6\% across seven stress and production-derived traces relative to a state-of-the-art KV-cache-based reactive autoscaler on the same runtime. These results show that, under shared GPU contention, effective LLM autoscaling requires cross-model comparison of current service deficit and coordinated capacity movement, rather than independent reactions to each model's local pressure.

{\footnotesize
\bibliographystyle{IEEEtran}
\balance
\bibliography{references}
}


\end{document}